\documentclass[final,5p,times,twocolumn]{elsarticle}

\usepackage{graphicx}
\usepackage{amssymb}
\usepackage{amsmath}
\usepackage{empheq} 
\usepackage{comment}
\usepackage{balance}
\usepackage{enumerate}
\usepackage[inline,shortlabels]{enumitem}
\usepackage[usenames,dvipsnames]{color}
\usepackage[colorlinks=true,linkcolor=NavyBlue,citecolor=NavyBlue,urlcolor=NavyBlue]{hyperref} 

\biboptions{sort&compress} 

\newcommand{\pd}[2]{ \frac{\partial #1}{\partial #2}} 
\newcommand{\mw}[1]{\left< #1 \right>} 
\newcommand{\nt}[3]{\int \limits_{#1}^{#2} \mathrm{d}#3~} 
\newcommand{\bb}[4]{#1_{#2, \hspace{0.3mm}#3}(#4)} 

\makeatletter
\def\@author#1{\g@addto@macro\elsauthors{\normalsize%
    \def\baselinestretch{1}%
    \upshape\authorsep#1\unskip\textsuperscript{%
      \ifx\@fnmark\@empty\else\unskip\sep\@fnmark\let\sep=,\fi
      \ifx\@corref\@empty\else\unskip\sep\@corref\let\sep=,\fi
      }%
    \def\authorsep{\unskip,\space}%
    \global\let\@fnmark\@empty
    \global\let\@corref\@empty  
    \global\let\sep\@empty}%
    \@eadauthor={#1}
}
\makeatother

\makeatletter
\def\ps@pprintTitle{%
 \let\@oddhead\@empty 
 \let\@evenhead\@empty
 \def\@oddfoot{}%
 \let\@evenfoot\@oddfoot}
 \makeatother

\journal{~}
\begin{document}
\begin{frontmatter}

\title{Prediction and Prevention of Disproportionally Dominant Agents in Complex Networks}

\author[add1,add2]{Sandro Claudio Lera}
\ead{slera@mit.edu}

\author[add1]{Alex `Sandy' Pentland}

\author[add2,add3,add4]{Didier Sornette}

\address[add1]{\scriptsize Massachusetts Institute of Technology, Massachusetts, USA}
\address[add2]{\scriptsize Southern University of Science and Technology, Shenzhen, China}
\address[add3]{\scriptsize ETH Zurich, Zurich, Switzerland}
\address[add4]{\scriptsize Tokyo Tech World Research Hub Initiative (WRHI), Tokyo, Japan}

\begin{abstract}
We develop an early warning system and subsequent optimal intervention policy to avoid the formation of disproportional dominance (`winner-takes-all') in growing complex networks.
This is modeled as a system of interacting agents, whereby the rate at which an agent establishes connections to others is proportional to its already existing number of connections and its intrinsic fitness.
We derive an exact 4-dimensional phase diagram that separates the growing system into two regimes:  one where the `fit-get-richer'  (FGR) and one where, eventually, the `winner-takes-all' (WTA).
By calibrating the system's parameters with maximum likelihood, its distance from the WTA regime can be monitored in real time. 
This is demonstrated by applying the theory to the eToro social trading platform where users mimic each others trades. 
If the system state is within or close to the WTA regime, we show how to efficiently control the system back into a more stable state along a geodesic path in the space of fitness distributions.
It turns out that the common measure of penalizing the most dominant agents does not solve sustainably the problem of drastic inequity.
Instead, interventions that first create a critical mass of high-fitness individuals followed by pushing the relatively low-fitness individuals upward is the best way to avoid swelling inequity and escalating fragility.
\end{abstract}

\end{frontmatter}


In just about a decade, a handful of companies have contributed to an alarmingly centralized world wide web. 
This hardly resembles the Internet's founding credo of net neutrality. 
Dominance on the Internet is not self-contained, but transcends the economy as a whole, such as digital advertising and e-commerce. 
Beyond economic concerns, such unprecedented digital monopolies have important implications for politics, and society \cite{Moore2018}. 
But how can one tell that a company is too large? 
At what point should regulators interfere and how? 
We present a framework to answer these and similar questions.  
At its heart is the law of preferential attachment, stating that the growth rate is proportional to size. 
We consider a system of interacting agents, whereby the rate at which any given agent establishes connections to others is proportional to its already existing number of connections and the agent's intrinsic fitness in attracting connections. 
This representation is known to carry a risk of centralization \cite{Bianconi2001}, in the sense that removal of a few dominating agents would collapse the entire system. 
We derive a systematic classification of the system's phase space and show under what circumstances such unfavorable centralization occurs. 
This allows us to anticipate the emergence of overly dominant agents ex ante and construct methods for early intervention.
In socio-economic systems, typical countermeasures against centralization are progressive taxes, anti-trust laws and similar legislation. 
However, our analysis reveals that such an approach may be ineffective because it addresses only the symptoms - disproportionally dominant agents - rather than the underlying cause - a fundamentally imbalanced system that catalyses such dominance. 
Instead of punishing the most competitive agents, one should foster more balanced growth and competition by improving the relative fitness of under-represented agents. 

The applications of our framework are manifold, since fitness-adjusted preferential attachment has been shown to approximate the growth dynamics of many systems at a phenomenological level. 
Examples include 
the evolution of the world-wide-web \cite{Kong2008}, 
citation networks \cite{Golosovsky2017}, 			
trust relationships \cite{Papadopoulos2012},  
social media \cite{Pham2016}, 
streaming services \cite{Pongnumkul2018}, 
photon emission rate \cite{Golosovsky2018}, 
supply chains \cite{Atalay2011},
Escherichia coli metabolic networks \cite{Papadopoulos2012}, 
market investments \cite{Garlaschelli2005},
and transactions on the bitcoin blockchain \cite{Aspembitova2019}. 
A theoretical justification for fitness-based behavioral preferential attachment has recently been provided as the result of strategic minimization of maximum exposure to least fit nodes \cite{Bell2017}.
In this article, we apply the method on the eToro trading platform, where participants observe others' trades and can chose to either engage in their own strategies or copy others.
We find that the platform does not give rise to``winners that take it all'', which we attribute to the inherent component of luck that underlies trading performance. 
We also discuss the implications of theses insights for a universal base-income, progressive tax, anti-trust laws and social-networks.

\section{Model Definition \& Regime Characterization} 
\label{sec:preferential_attachment}

In this section, we propose an exact criterion specifying the conditions for which the dominance of an agent (firm, city, website, individual, etc.) becomes overwhelming. 

\subsection{Fitness-Based Preferential Attachment \& Deletion}

Aiming at a generic description, preferential attachment is a natural first building block \cite{Simon1955,BookSaiMalSor09,Barabasi1999}. 
Assuming some heterogeneity among agents,
it is appropriate to add to it a fitness-based proportional attachment \cite{Bianconi2001a}, which has been shown \cite{Bell2017} to be particularly suitable to describe growth dynamics in complex networks. 
In practice, the appropriateness of this form may be tested by means of specific Bayesian statistical methods \cite{Pham2016}.

We thus consider an undirected network that is growing by attaching one agent and $m$ edges per unit time.
Edges may also be established between already existing nodes.
Each agent $a_i$ has an intrinsic `attractiveness' or, from hereon, `fitness' $\eta_i \in (0,1)$ sampled from some fitnesses distribution $\rho$. 
Agent $a_i$'s number of connections, i.e. its degree, is denoted by $k_i$. 
At each time-step, agent $a_i$ establishes a new connection with probability $p_i$ proportional to the product of its fitness and degree, $p_i = \eta_i ~ k_i / \sum_j ( \eta_j p_j )$. 
The fitter and the more connected the agent, the more attractive it is. 
To add one more realistic, yet generic assumption \cite{Kong2008a}, we allow agents to be removed from the system (failure, death, etc.) with probability proportional to $c \eta_i^\omega$.
Here, $c \geqslant 0$ is the base rate of removal and $\eta_i^\omega$ accounts for the dependence of the probability of failure on the fitness of the agent. 
For $\omega > 0$, the fit agents are more likely to fail, for instance  due to targeted attacks. 
For $\omega < 0$, weak agents are more likely to fail, i.e. $\eta$ itself is a measure of robustness with respect to failure. 

Assuming that the $i$-th agent is still alive at time $t$, the mean field rate at which its degree $k_i$ changes reads
\begin{equation}
	\pd{k_i}{t} = m \frac{ \eta_i k_i}{S(t)}  - \frac{c}{1-c}~ \frac{ \mw{\eta}^\omega }{ \mw{ \eta^\omega }}  \frac{k_i}{t}
	\label{eq:micro_equations_formal}
\end{equation}
where $S$ is a normalization factor and expectation values $\left< \cdot \right>$ are taken over the fitness distribution $\rho$. 
The first term in the right hand side of  \eqref{eq:micro_equations_formal} is a direct consequence of the attachment rule described above.
The second factor is a product of two independent probabilities: 
the probability $\sim c \mw{\eta}^\omega$ that an agent is deleted, and the probability $\sim k_i / t$ that agent $a_i$ is connected to that failed agent. 
A detailed derivation of this equation and subsequent results is found in the supplementary information (SI).
\footnote{The SI Appendix is available from the authors upon request.}

While  \eqref{eq:micro_equations_formal} captures the growth dynamics of a large class of systems at least to first order, 
it may also be extended to take into consideration for instance 
preferential deletion \cite{Kong2008a}, 
sub-linear or super-linear preferential attachment \cite{Krapivsky2002}, 
assortative mixing \cite{Newman2002}, 
multi-dimensional fitness \cite{Fowler2009}, 
or different boundary conditions such as minimal lower-bounds or inflow of new agents at different rates \cite{BookSaiMalSor09}.

\subsection{The fit-get-richer (FGR) and the winner-takes-all (WTA) regimes}

\begin{figure*}[!htb]
	\centering
	\includegraphics[width=\textwidth]{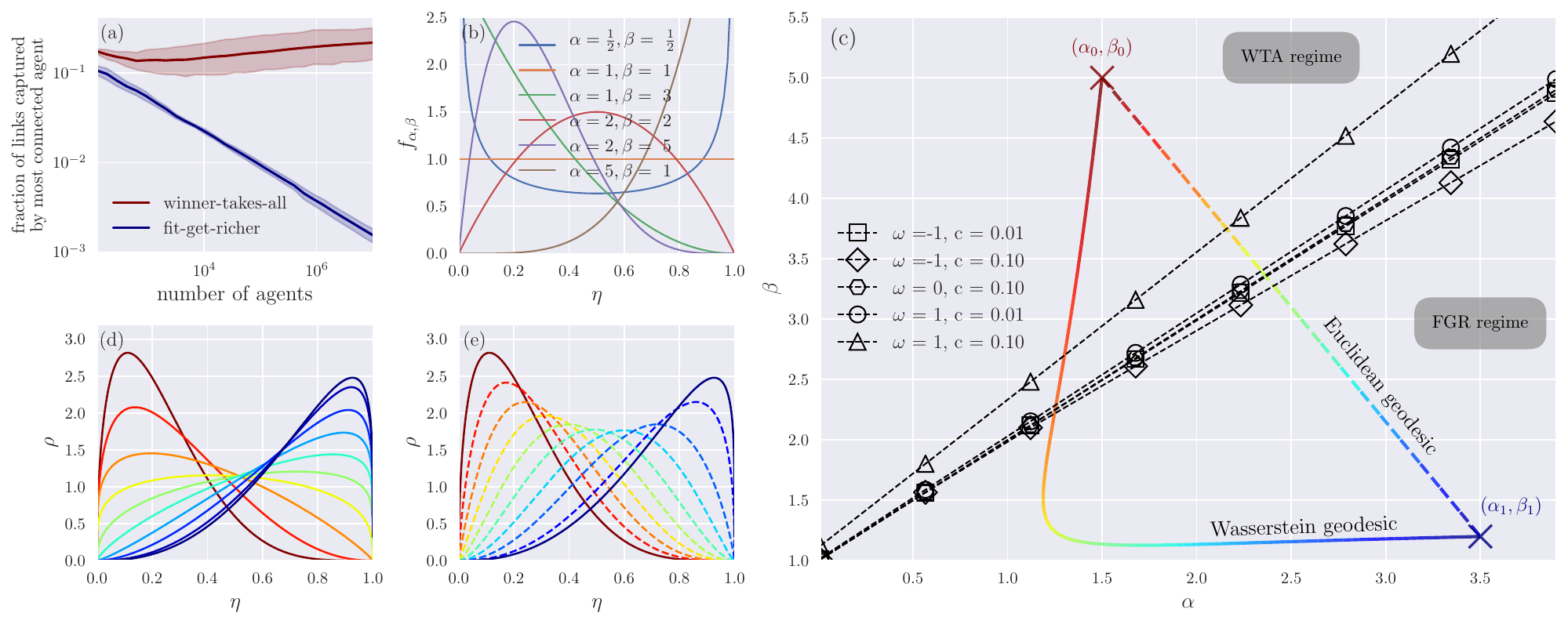} 
	\caption{	(a)
			Fraction of links connected to the most dominant agent $n_{\rm DK}$ in the system, as a function of the total number of agents. 
			The mean values and $\pm$ one standard deviation bands are obtained over a sample of $100$ realizations.
			We have fixed $c=1\%$ and $\omega=0$. 
			Two different fitness distributions are used, one in the WTA and one in the FGR regime (red and blue cross in panel (c)).
			(b)
			Beta distribution density for different parameters $\alpha$ and $\beta$ used to parametrize the class of fitness distributions on $[0,1]$. 
			As $\alpha$ and $\beta$ vary, all typically relevant shapes of the fitness distribution are sampled. 
			The WTA regime is  characterized by the bulk of the probability mass being concentrated on agents of relatively low fitness, with only few fit agents that end up dominating the system (e.g. $\alpha=2, \beta=5$). 
			(c) 
			Phase diagram showing the domains of existence for the fit-get-richer (FGR) and the winner-takes-all (WTA) regimes, for different preferential deletion parameters $c$ and $\omega$ in the space of the parameters $(\alpha,\beta)$ 
			of the Beta-fitness distribution. 		
			The WTA regime corresponds to values where $I^* < 1$ (\eqref{eq:phase_transition_condition}). 
			This WTA regimes lies above a line parametrized by $(c,\omega)$.   
			(d) Geodesic with respect to Wasserstein-2 metric, i.e. cost minimizing interpolation from a  Beta distribution in the WTA regime to a Beta distribution in the FGR regime. 
				The (approximate) path in the space of Beta-distributions is shown in panel (c). 
			(e) Distributions along the shortest path with respect to the Euclidean distance in the $(\alpha,\beta)$-plane (cf. dashed line in panel (c)). 
				This intervention policy is more expensive than the one along the Wasserstein geodesic.
			} 
	\label{fig:phase_diagram}
\end{figure*}

As shown in the SI,  \eqref{eq:micro_equations_formal} gives rise to two distinct asymptotic regimes: 
the fit-get-richer (FGR) and the winner-takes-all (WTA). 
In the later, the system is largely dependent and controlled by just a few agents. 
This regime is analogous to a Bose-Einstein condensate in statistical quantum mechanics \cite{Bianconi2001,YukaSor12}
and the dominant agents are known as `dragon-kings' in socio-economic \cite{Sornette2012a} and complex systems \cite{Sornette2009}.
The notion of disproportional dominance manifests itself in the number of connections of those agents (representing their influence/importance/wealth/\ldots). 
This is made rigorous as follows: 
there exists at least one agent $a_{DK}$ whose ratio of its degree $k_{DK}$ to the total number of connections does not decay to zero in the limit of infinite system size. 
In other words, even as the system grows to infinity, a non-zero fraction of connections is controlled by $a_{DK}$.
The influence of such an agent is felt in the entire system no matter what the total size of the network (Figure \ref{fig:phase_diagram}(a)). 
From a stability point of view, the system is then strongly reliant on a few agents, and might collapse if those fail. 
From an economic point of view, any action of such  an agent affects the entire economy, reminiscent of the companies mentioned in the introduction. 

The WTA regime occurs if and only if 
\begin{equation}
	I^*
	\equiv 
	\nt{0}{1}{\eta}
	 \frac{ \rho(\eta) }{ 
	  			\frac{ \xi }{ \xi^\omega + \mw{\eta}^\omega + \frac{1-c}{c} \mw{\eta^\omega} }  \left( \frac{ \mw{\eta}^\eta + \frac{1-c}{c} \mw{\eta^\omega} }{\eta} + \frac{1}{\eta^{1-\omega}} \right)  - 1
				} 
				< 1
	\label{eq:phase_transition_condition}
\end{equation}
where $\xi$ is a constant that depends on $\rho, c$ and $\omega$, but is typically close to $1$. 
Expectations $\left< \cdot \right>$ are calculated with respect to the distribution $\rho$  from which $\eta$ is sampled. 
Importantly, \eqref{eq:phase_transition_condition} depends on the fitness landscape $\rho$ as a whole, and not just on the fitness of a few agents.
This suggests that acting on the dominant agents by removing many of their connections or targeting their growth specifically will not be effective: 
this may transiently decrease their influence but is short-lived as the growth dynamics leads to the resurgence of new agents that dominate.
The problem is systemic: 
as we shall see, in order to prevent the occurrence of unwanted winners-takes-all agents, one should act on the fitness landscape $\rho$ as a whole.

Finally, it is  important to note that the FGR regime is not to be mistaken for a state of `equality'. 
The asymptotic size distribution is still heavy-tailed with a power law exponent controlled by the shape of $\rho$ \cite{Bianconi2001a}. 
However, the important difference is that the influence of any given agent is more localized and negligible within a large enough system, 
in contrast to the WTA regime where some agents dominate system-wide. 

\subsection{Parametric Regime Classification} 
\label{sec:regimes}

\eqref{eq:phase_transition_condition} shows that the appearance of WTA agents depends on the deletion parameters $c$ and $\omega$, as well as the entire distribution of fitnesses $\rho$. 
In order to classify the possible different regimes, we propose to explore the space of fitness distributions parameterized by the Beta-distribution whose density is given by 
$\bb{f}{\alpha}{\beta}{\eta}  \propto  \eta^{\alpha-1} ~ 	(1-\eta)^{\beta-1}$ where varying $\alpha, \beta > 0$ samples all shapes of practical interest, see Figure \ref{fig:phase_diagram}(b). 

The winner-takes-all (WTA) condition \eqref{eq:phase_transition_condition} can now be expressed as a function of four parameters $\{\alpha, \beta, \omega, c\}$, via the function $I^*(\alpha, \beta, \omega, c)$ and the two corresponding regimes are FGR ($I^* > 1$) and WTA $(I^* < 1)$. 
In general, the integration in \eqref{eq:phase_transition_condition} with $\rho$ has to be solved numerically.  
Figure \ref{fig:phase_diagram}(c) shows the domain of existence for the WTA and FGR regimes, for different preferential deletion parameters $c$ and $\omega$ in the space of the parameters $(\alpha,\beta)$. 
Comparing with Figure \ref{fig:phase_diagram}(b), we see that the WTA regime corresponds to fitness landscapes in which the bulk of the agents have relatively low fitness.
Smaller values of $\omega$ favor the failure of relatively low fitness agents, and hence increase the area of the WTA regime.
In contrast, for large $\omega$, the fit-get-richer (FGR) regime increases in importance, as more fit agents tend to be removed by targeted attacks. 
The larger $c$ is, the stronger is this effect.

\section{Empirical Calibration on the eToro Social Trading Platform}
\label{sec:eToro}

\begin{figure}[!htb]
	\centering
	\includegraphics[width=0.5\textwidth]{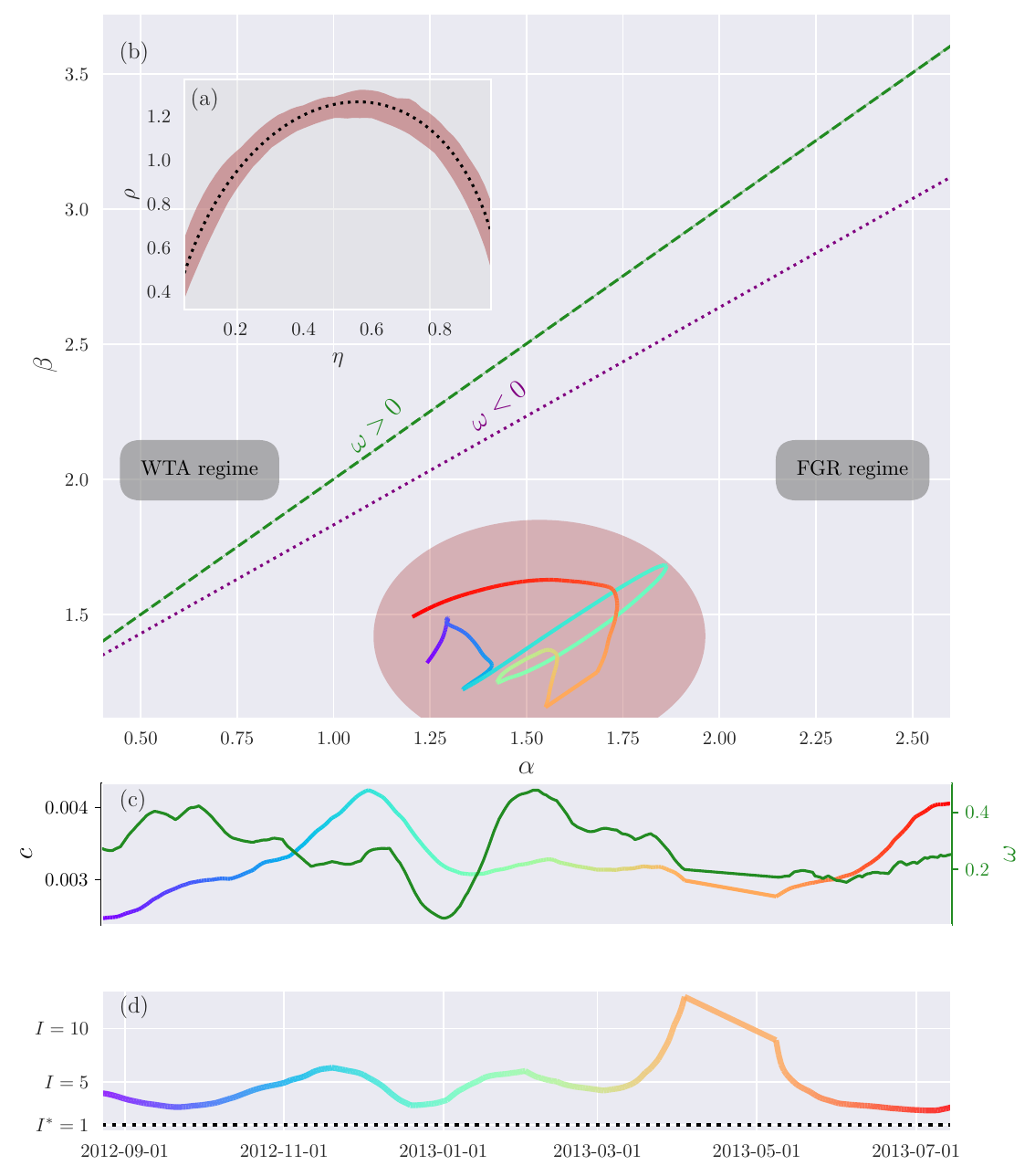}
	\caption{	Analysis of eToro trading network at different times $t$ from July 2012 to November 2013. 
			(a): 
			Time-averaged distribution of trader fitnesses $\hat{\rho}$ (dashed black line) as well as $80\%$ confidence interval as determined over all times $t$. 
			(b) 			
			Evolution of the eToro network in the $(\alpha, \beta)$ plane obtained by the procedure explained in the text.
			The straight lines separate the WTA regime from the FGR regime.
			The green, dashed line corresponds to the regime separation with time-averaged $\hat{\omega} \approx 0.3$ and $\hat{c} \approx 0.003$ value. 
			The purple, dotted  line highlights the hypothetical effect of removing relatively low fitness traders at a high rate ($\omega = -0.3, c = 0.5$). 
			(c) 
			Time evolution of parameters $c$ (multi colors an left axis) and $\omega$ (green and right axis). 
			(d) 
			Time evolution of the parameter $I^*$ compared with the horizontal line at $I^*=1$ that separates
			the two regimes (FGR for $I^* > 1$ and WTA for $I^* < 1$). 
			}
	\label{fig:etoro}
\end{figure}

We apply the above theory to the social trading dynamics on the multi-asset brokerage platform eToro. 
On eToro, traders have access to different assets, in particular foreign exchange markets, and can leverage their position up to 400 times for both long and short positions. 
The feature that most distinguishes eToro from other trading platforms is its `OpenBook' social investment function. 
Instead of making their own trading decisions, users can chose to `mimic' one or several other traders and automatically execute the same trades as they do. 
Since more popular traders are more prominently displayed on the platform, they are more likely to be noticed and hence followed. 
This suggests that a preferential attachment model is suitable to model the imitation tendency.
The decision to follow a specific trader also depends on idiosyncratic, heterogeneous properties (e.g. recent performance or risk level), thus adding a fitness component seems appropriate. 
To confirm the applicability of fitness-adjusted preferential attachment, 
we have applied a Bayesian statistical method for the joint estimation of preferential attachment and node fitness without imposing functional constraints \cite{Pham2016}. 
See SI for details. 

Translating to our model, each trader is an agent, and the number of traders followings a given trader $a_i$ defines its number of connections (degree) $k_i$. 
The rate at which a trader attracts new connections is a measure of the attachment probability $p_i$. 
In this way, by measuring agent $a_i$'s popularity $k_i$ as well as its current rate of attraction $p_i$, one can infer an effective fitness $\eta_i \propto p_i / k_i$ without detailed knowledge of the underlying system dynamics. 
The access to all the active accounts and traders allows us a faithful reconstruction of the underlying network. 
We measure $p_i$ and $k_i$ over rolling windows at different times $t$ from July 2012 to November 2013, such that the evolution of the fitness can be tracked: $t \mapsto \eta_i(t)$.
After having inferred the fitness of every agent at time $t$, we construct an empirical fitness distribution $\hat{\rho}(t)$ as a histogram over all fitnesses. 
The distribution of fitnesses turns out to be fairly stationary and unimodal, with most traders having a fitness around $\eta_i \approx 0.4$ (Figure \ref{fig:etoro}(a)). 
The histogram $\hat{\rho}(t)$ is fitted fitted by the Beta-distribution via maximum likelihood to obtain the parameters $\hat{\alpha}(t)$ and $\hat{\beta}(t)$.  
In accordance with the stationarity of $\hat{\rho}$, the evolution of $t \mapsto \left( \hat{\alpha}(t), \hat{\beta}(t) \right)$ is confined to a small radius around $\hat{\alpha} \approx 1.25$ and $\hat{\beta} \approx 1.25$  (Figure \ref{fig:etoro}(b)).
Similarly, the deletion parameters $\hat{c}(t)$ and $\hat{\omega}(t)$ are inferred from direct observation at different times $t$ and are found to be
stationary and fluctuating within limiting bounds (Figure \ref{fig:etoro}(c)).
See SI for details on methods. 
Here, we have defined node removal as the process whereby a trader is no more being mimicked by anyone after having been mimicked previously.
Interestingly, $\hat{\omega}$ is consistently positive, suggesting that particularly fit traders are more likely to disappear. 
Our method is phenomenological in that it captures idiosyncratic, system specific details into the fitness term. 
To interpret this result, one must thus apply domain-specific knowledge. 
On eToro, a trader's recent performance is displayed to others by means of several metrics (percentage gain, portfolio volatility etc.) 
We can then expect $\eta$ to be a (generally complex) function of these variables.
Performance on financial markets is known to have a large contribution of luck \cite{Fama2010}
and presumably even more so for the primarily retail-traders that are active on eToro. 

In the SI, we show that the most mimicked (i.e. high $\eta$) traders indeed correspond to the ones who seem to outperform the market on short time scales, but that this performance is not sustained over longer times. 
Since, on short time-scales, luck is easily mistaken as skill \cite{SorWheatCau19}, the mimicking functionality of eToro gives rise to imitation of noise traders. 
Implied over-confidence may then be the reason that more risk is being taken, resulting in large losses and explaining why particularly fit traders are more likely to vanish from the platform. 

These insights highlight a secondary benefit of our approach: it complements and guides a more system specific understanding of the problem at hand. 
As long as the system allows for a representation of the form \eqref{eq:micro_equations_formal} (or a related variant thereof),  one can absorb the intrinsic system dynamics into $\eta$. 
The phenomenological model then allows one  to determine the state of the system. 
System specific knowledge is again required to understand how the fitness distribution is to be modified in practice (cf. also below).

We now continue with the main purpose of the method: assessing the state of disproportional dominance.
Given the estimates of $\alpha, \beta, c$ and $\omega$, we can now calculate $t \mapsto I^* \left( \hat{\alpha}, \hat{\beta}, \hat{\omega}, \hat{c} \right)$ from \eqref{eq:phase_transition_condition} and track its distance from the transition threshold $I^* = 1$. 
For the case of eToro, we find that the system dynamics is confined in the stable FGR regime $I^* > 1$ (Figure \ref{fig:etoro}(d)), suggesting that the competitive environment is enough to avoid the emergence of disproportionally dominant traders. 
This is alternatively visualized in the $\left(\hat{\alpha}, \hat{\beta} \right)$-plane (Figure \ref{fig:etoro}(b)), where we can see that the dynamics is well below the separation line to the WTA regime. 
If the dynamics was such that unfit traders are more likely to exit ($\omega < 0$) at higher rates $(1 > c \gg 0$), the system would however lie closer to the critical threshold at $I^* = 1$.  

We note that the strength of the above methodology is that it can also be implemented to incomplete or noisy data, as the regime classification depends only on the empirical distribution of fitnesses $\hat{\rho}$ as a whole.
A current limitation is that the fitnesses are assumed stationary, or at least that they change over time-scales that are larger than for changes in the number of established connections. 
The SI of this article confirms this is the case for eToro.
Relaxing this assumption is part of future research. 
Realizing that a large part of the traders' performance observed on eToro is likely driven by luck, 
it seems particularly interesting to take into consideration fitness values that decay with time  \cite{Medo2011}.

\section{Optimal Control of Disproportional Dominance} 
\label{sec:prevention}

The analysis of theToro social trading network in the previous section showed an instance where the network was confined in the FGR regime. 
In less competitive, large-scale economic systems, this may not always be the case. 
In such situations, an external regulator (e.g. the state) may wish to intervene to ensure greater opportunities for new entrants and less inequity. 
Interventions only targeting the most dominant agents may appear `unfair', as it punishes arguably highly fit agents. 
Perhaps more important, such intervention is inefficient in light of \eqref{eq:phase_transition_condition}, which shows that the emergence of the WTA regime is a consequence of the distribution of fitnesses as a whole. 
Rather than just acting on the right tail of the distribution in form of punishments, more holistic, distribution-wide interventions are called for.
In particular,  as inferred from panel (b) and  (c) in Figure  \ref{fig:phase_diagram}, it is the high population of weak agents that needs to be addressed. 

Relying again on our parametrization of the fitness space in terms of the Beta-distribution, we formulate the regulator's intervention as an action that aims to modify the distribution of fitnesses.
Mathematically, this amounts to the problem of shifting optimally the distribution from an initial shape characterized by $(\alpha_0, \beta_0)$ for instance inside or uncomfortably close to the WTA regime to another shape represented by $(\alpha_1, \beta_1)$ inside the FGR domain (see Figure \ref{fig:phase_diagram}(c)).  
Practically, this amounts to increasing the concentration of high-fitness agents relative to relatively low-fitness agents (e.g. in form of subsidies, start-up promotion, 
professional education and so forth).
Technically, we formulate the intervention technique as an optimal transport problem, seeking the most cost efficient way of transforming one distribution into another.
We assume that increasing an agent's fitness from $\eta_1$ to $\eta_2$ is associated with a cost $c(\eta_1, \eta_2)$. 
The problem then amounts to finding a transport map $\tau: [0,1] \to [0,1]$ that pushes $\bb{f}{\alpha_0}{\beta_0}{\cdot}$ onto $\bb{f}{\alpha_1}{\beta_1}{\cdot}$ while minimizing the total cost $\nt{0}{1}{\eta} c(\eta, \tau(\eta)) ~\bb{f}{\alpha_0}{\beta_0}{\eta}$.
The solution of the problem depends on the functional shape of $c(\cdot, \cdot)$. 
Here, we provide a specific example by assuming a quadratic cost $c(\eta_1, \eta_2) = \left| \eta_1 - \eta_2 \right|^2$. 
This is a reasonable assumption for systems where fitness can be controlled directly or indirectly via external manipulation.
Whether a quadratic cost is appropriate generally depends on the specific system under scrutiny, but our method generalizes to other functional shapes for $c$. 
In full generality, different cost functions will yield different intervention protocols. 
However, as discussed in the SI Appendix, our conclusions remain similar when extended to $c(\eta_1, \eta_2) = \left| \eta_1 - \eta_2 \right|^\delta$ for any $\delta \geqslant 1$. 
By varying $\delta$, a large class of cost functions may be approximated reasonably well.

Denote by $\bb{F}{\alpha}{\beta}{\cdot}$ the cumulative distribution function of the Beta distribution, and by $\bb{Q}{\alpha}{\beta}{\cdot} \equiv \bb{F^{-1}}{\alpha}{\beta}{\cdot}$ its inverse (quantile) function.
It can be shown \cite{Peyre2019} that the cost minimizing transport map $\tau$ is of the form $\tau(\eta) = \bb{Q}{\alpha_1}{\beta_1}{ \bb{F}{\alpha_0}{\beta_0}{\eta} }$.
The interpolation from $\bb{f}{\alpha_0}{\beta_0}{\cdot}$ to $\bb{f}{\alpha_1}{\beta_1}{\cdot}$ is equivalent to a geodesic path with respect to the Wasserstein-2 metric. 
It is parametrized by $f_t(\eta) = \pd{\tau_t}{\eta} ~ \bb{f}{\alpha_1}{\beta_1}{ \tau_t(\eta) }$ with $\tau_t =  t~ \eta + (1-t)~\tau(\eta)$ and $t$ is running from $0$ to $1$.
Hence, $f_t(\eta)$ interpolates from $f_0(\eta) = \bb{f}{\alpha_0}{\beta_0}{\eta}$ to $f_1(\eta) = \bb{f}{\alpha_1}{\beta_1}{\eta}$ through the space of fitness distributions along a geodesic path with respect to the Wasserstein-2 metric
(Figure \ref{fig:phase_diagram}(d)). 
We can again approximate any intermediate distributions $f_t(\eta)$ as Beta-distribution with parameters $(\alpha_t, \beta_t)$ (see SI for mathematical details). 
This traces out a path $t \mapsto (\alpha_t, \beta_t)$ that can be interpreted as the most cost-efficient intervention plan. 
In Figure \ref{fig:phase_diagram}(c), such a geodesic path is shown, starting from the red cross deep in the WTA domain and ending on the blue cross in the FGR domain.
Panel (d) of Figure \ref{fig:phase_diagram} presents a sequence of snapshots for the distribution of fitnesses, as it changes along the optimal path. 
This geodesic in the distribution space deviates very strongly from the naive solution of taking a straight path in the 2-dimensional $(\alpha,\beta)$ plane (which would be the geodesic with Euclidean distances). 
As already mentioned, the WTA regime corresponds to a relatively low-fitness population with a few high fitness individuals, who eventually become winners that `take it all'. 
The most cost efficient intervention (Wasserstein geodesic) focuses first on the high fitness part of the distribution, by decreasing $\beta$ which describes the part of the fitness population close to $1$ while keeping $\alpha$ approximately constant. 
Then, the intervention shifts to increasing $\alpha$ at quasi-fixed $\beta$, thus working on the part of the fitness population close to $0$.
In contrast, the straight line (Euclidean geodesic) works on both $\alpha$ and $\beta$ simultaneously, which corresponds to first building a large pool of agents with intermediate fitnesses (Figure \ref{fig:phase_diagram}(e)). 
The optimal intervention thus amounts to  decreasing the number of agents of relatively low fitness while increasing the number of relative fit ones,
while avoiding a population that peaks at intermediate fitness levels.

\section{Implications \& Applications} 
\label{sec:applications}

Since the growth dynamics of many systems may be captured at least to first order by fitness-based preferential attachment 
\cite{Kong2008,Golosovsky2017,Papadopoulos2012,Pham2016,Golosovsky2018,Atalay2011,Papadopoulos2012,Pongnumkul2018,Garlaschelli2005,Aspembitova2019}, 
the insights from this article can be applied to a large class of systems. 
Let us discuss a few examples. 

Already in the early 20th century, Gibrat had recognized that the growth dynamics of firms may be described with proportional growth  \cite{Gibrat1931}. 
This model has ever since been extended numerous times to take into consideration effects such as 
firm heterogeneity \cite{Luttmer2011,Axtell2008,Wyart2003,Takayasu2014},
minimum firm size \cite{Gabaix1999}, 
births and deaths of firms \cite{Grossman1993,Steindl1965,Luttmer2007},
merger and acquisitions \cite{Lera2017}, 
or bankruptcy \cite{Saichev2010a}.
In the context of our model, the firm dynamics can be described for instance via preferential attachment of the supply-chain network \cite{Atalay2011}.
The number of customers of a firm is denoted by $k$ and $\eta$ is the rate at which its customer base grows. 
Our results then have implications for instance for the formulation of anti-trust laws. 
Regulators are concerned with the merger of two large firms. 
The impact of a merger is often assessed in terms of concentration of market power \cite{Kamien1990}. 
Firm size may be one important aspect, especially in sectors with only a few firms.
However, our results suggest that a potentially greater harm stems from the less monitored acquisitions of successful smaller firms (such as start-ups) that gradually deplete the pool of high fitness agents. 

At the level of personal wealth, the concept of fitness based preferential attachment is also relevant. 
The richer an agent, the more likely they are to attract more wealth by being exposed to more opportunities and transactions \cite{Garlaschelli2005,Earl2010,Aspembitova2019}. 
Denote by $k$ the wealth of an agent.
Then, it holds $\mathrm{d} k \propto \eta k$ where the fitness $\eta$ is the growth rate in this context. 
From such a representation, the effect of a universal base income is to put a floor on the minimum wealth $k_i$ that any agent may have. 
It thereby acts as a reflecting lower boundary and this boundary condition can even make the wealth distribution more unequal, transforming it from log-normal to power law \cite{Gabaix1999}. 
However, to sustainably address the problem of income and wealth inequality, it is the growth rates (i.e. fitnesses) themselves that need to be addressed. 

While our model sheds light on what should be done and where resources should be targeted, system specific domain knowledge is required to put these insights into best practice. 
A progressive tax system that takes from the rich and gives to the poor may free up the necessary resources, but does not solve the problem sustainably by itself. 
While it is expected that the ability to generate income (i.e. fitness) changes itself with increased wealth, the uncontrolled, raw transfer of money may be inefficient, short-lived or localized \cite{Kuhn2011,Ager2019}. 
Instead, system specific actions should be designed with the goal of increasing growth opportunities for low-income individuals e.g. by enabling them to save money and earn returns, or better access to opportunities and investments.

Streaming services and online vendors often recommend new products to its users based on their popularity (degree $k$) and rating (fitness $\eta$) \cite{Pongnumkul2018}. 
To additionally capture a component of personalized recommendation and the bi-partite nature of the product-customer space, \eqref{eq:micro_equations_formal} could be extended to take some users' characteristics into consideration.
In light of our results, to avoid that a few products completely dominate and to ensure the offering of a large variety of large fitness products, sufficient resources obtained from sales of successful products should be directed
towards supporting creativity and inventions of new styles.

\section{Concluding Remarks} 
\label{sec:conclusion}

The above results have shown that the emergence of disproportional dominance is not explained as the action of a single agent, but rather a wholistic trait of the system. 
Our model highlights the importance of vigorous holistic interventions, which work on the full distribution of fitnesses, and focus on developing a high-fitness population. 
In particular, interventions should emphasize the support of relatively weak agents rather than punishing the most dominant ones. 
This may seem counterintuitive, but the sequential actions of first building a strong base of high-fitness individuals followed by pushing upward the fitnesses of the relatively low-fitness individuals is a more effective way of reducing inequity and increasing robustness.

Given the generic set-up which may be adapted to different boundary conditions and its robustness to noise, 
our methodology has the potential of becoming a standard tool for dynamic risk monitoring in interacting systems ranging from small scale trading platforms up to globally interconnected social and economic systems.

\balance
\bibliographystyle{unsrt}
\bibliography{bibliography}

\end{document}